# New Sol-Gel Synthesis of Ordered Nanostructured Doped ZnO Films

N. R. S. Farley[a)], C. R. Staddon, L. X. Zhao, K. W. Edmonds, B. L. Gallagher
*School of Physics and Astronomy, University of Nottingham, Nottingham NG7 2RD, UK*
D. H. Gregory
*School of Chemistry, University of Nottingham, Nottingham NG7 2RD, UK*

A novel sol-gel route to c-axis orientated undoped and Co, Fe, Mn and V doped ZnO films is reported. Sols were prepared from a hydrated zinc acetate precursor and dimethyl formamide (DMF) solvent. Films were spin-coated on to hydrophilic sapphire substrates then dried, annealed and post-annealed, producing almost purely uniaxial ZnO crystallites and a high degree of long-range structural order. Specific orientation of hexagonal crystallites is demonstrated both perpendicular and parallel to the substrate surface. Cobalt doping resulted in the formation of columnar ZnO nanocrystals. Vanadium doped films formed the spinel oxide $ZnAl_2O_4$, resulting from the reaction between ZnO and the sapphire substrate. Structural, optical and morphological characterisation demonstrated the high quality of the films.

## INTRODUCTION

ZnO has a long history of usage for pigments and protective coatings on metals. The electrical, optoelectronic and photochemical properties of undoped ZnO has resulted in use for solar cells [1,2], transparent electrodes [2-4] and blue/UV light emitting devices [5]. Ordered c-axis orientation of ZnO crystallites perpendicular to the substrate surface is desirable for applications where crystallographic anisotropy is a prerequisite *e.g.* piezoelectric surface acoustic wave or acousto-optic devices [6]. Controlling the electrical properties of ZnO by transition metal doping can optimise material performance, and expands technological applications to ferromagnetic semiconductors [7] (spintronic materials). The long-range ferromagnetic exchange interaction between widely spaced dopant atoms demands a material with high structural quality, as demonstrated for III-V ferromagnetic semiconductors such as $Ga_{1-x}Mn_xAs$ [8-10]. Films with such a high degree of structural order are usually deposited epitaxially by molecular beam epitaxy (MBE) [5], pulsed laser deposition, chemical vapour deposition [6] or sputtering. Sol-gel methods are being employed increasingly for the low-cost fabrication of ordered high-specification materials since structural and morphological characteristics may be tuned in order to tailor the optical, electrical or magnetic properties of the material. Sol-gel methods may also be favoured since it is possible to produce a large number of samples rapidly and at a fraction of the cost of MBE. Sol-gel preparations are therefore ideal for exploratory studies for which large numbers of candidate materials, compositions or preparative conditions require screening.

In this study we describe a simplified sol preparation compared to those reported in the literature. Alcoholic solvents are often used to prepare sols; for instance, most sol-gel syntheses of undoped and doped c-axis orientated ZnO use solvents such as methoxyethanol [11-13] or ethanol/propanol [14,15]. Alcohols are commonly used because the alcohol also acts as a reagent. However, the solvent does not participate in the reaction forming ZnO from zinc acetate. Also, the solubility of zinc acetate in alcohols is quite low, so the Zn precursor is usually dissolved by heating the mixture to 60-100 °C. These methods also use additives such as mono/di/triethylamine [4,11-14,16-17] or lactic acid [15] to improve the stability and homogeneity of the sol by preventing uncontrolled hydrolysis and precipitation of ZnO.

---

[a)] Author to whom correspondence should be addressed; email: Nicola.Farley@nottingham.ac.uk

The procedure reported here is simpler than established methods since the sols consist only of two or three components: zinc acetate precursor, a dopant salt and the solvent. The low volatility of the solvent used in this study (dimethyl formamide, DMF) prevents premature and uneven drying, which may cause cracking and disrupt uniaxial crystallisation of the films. Zinc acetate also dissolves readily in DMF, so it is not necessary to heat the starting mixture. An additive is not required to improve sol stability and homogeneity. The process of sol formation is therefore simplified, and room temperature preparation reduces the risk of inadvertently altering the oxidation state of dopant ions prior to their incorporation into the oxide host.

We present structural, morphological and optical characterisation of $ZnO$ / $ZnAl_2O_4$ films, both undoped and doped with 6 mol % $Co^{2+}$, $Fe^{3+}$, $Mn^{2+}$, $Mn^{3+}$ or $V^{3+}$ with respect to Zn. Henceforth, the initial oxidation state of the dopants will be used to distinguish the samples.

**EXPERIMENT**

Each sol was synthesised from a zinc acetate $Zn(CH_3CO_2)_2 \cdot 2H_2O$ precursor (Fluka, 99.5+ %) and dimethyl formamide solvent (Fisher, >99 %). Dopant ions were provided by transition metal chlorides or acetates: $CoCl_2 \cdot 5\text{-}6H_2O$ (99.999 %, Aldrich), $FeCl_3 \cdot 6H_2O$ (97+ %, East Anglia Chemicals), $Mn(CH_3CO_2)_2 \cdot 4H_2O$ (99+ %, Aldrich), $Mn(CH_3CO_2)_3 \cdot 2H_2O$ (97 %, Aldrich) and $VCl_3$ (Aldrich). Zinc acetate and dopant salts were dissolved completely in DMF by stirring at room temperature. Six 30 mL, 0.6 M sols were prepared: undoped (transparent), $Fe^{3+}$ doped (red), $Co^{2+}$ doped (pink), $Mn^{2+}$ doped (transparent), $Mn^{3+}$ doped (dark brown) and $V^{3+}$ doped (dark green). The sols were stable and homogeneous; no particulates or precipitates were visible to the eye and their appearance was unchanged for several months. Sols were usually stirred overnight at room temperature before use.

Sapphire (0001) substrates (Union Carbide, USA) were cleaned and hydrophilically terminated by immersion in 1:1 concentrated $H_2SO_4$ : 30 % vol. $H_2O_2$ (piranha solution) for 15 min, followed by rinsing with deionised water and drying in air. Films were deposited by dropping sol onto the substrate using a 1 mL syringe fitted with a polypropylene filter with 0.2 μm pores (Whatman, BDH). Filtration was a necessary precaution to ensure that dust particulates were not incorporated into the films, since they are thought to disrupt uniaxial crystallisation [12]. As an additional precaution, the magnetic PTFE-covered stirrer bead also served to withhold any solid magnetic impurities that may have been present in the sol. Each film consisted of 5 layers, each spun at 4000 rpm for 30 s. After deposition, each layer was left to hydrolyse in humid air at room temperature for 1 h then dried at 300 °C for 10 min in an alumina crucible before spinning the next layer. Each spun layer must be thin enough to ensure the simultaneous evaporation of all solvent, thus preventing the cracking and disruption to crystallisation caused by gradual evaporation [11]. Each complete five layer film was annealed at 550 °C for 5 h, warming at 1 °C min$^{-1}$ in air, then post-annealed at 15 °C min$^{-1}$ to 850 °C for 5 h, cooling naturally. This annealing regime is similar to that reported by Wessler et al.[12] for undoped $ZnO$ / $ZnAl_2O_4$ with methoxyethanol solvent. All undoped and doped films appeared smooth, uniform and transparent.

Structural characterisation was performed using a Philips X'pert materials research X-ray diffractometer with a Cu anode and a β-filtering parafocusing mirror, generating Kα radiation at 1.542 Å and operating at 45 kV and 40 mA. All measurements were performed using a parallel plate detector and the beam size was 20 mm$^2$. 2θ/ω scans (analogous to θ/2θ scans) were recorded to determine the crystallinity and crystalline orientation of the films. The grazing incidence geometry was used for 2θ and ϕ scans in order to access specific out-of-plane, high symmetry reflections. The Scherrer formula was used to estimate the ZnO particle size from the 0002 peak width. Surface morphology was investigated using a Digital

Instruments multimode scanning probe microscope for tapping mode atomic force microscopy (AFM), utilising Al-covered single Si cantilevers (Ultrasharp, MikroMasch) and operating at about 150 kHz. Images were processed using the program WSxM (Nanotec Electronica, Spain). Optical qualities were assessed by photoluminescence measurements, utilising a Cd-He laser with an excitation wavelength of 325 nm and a beam diameter of about 1 mm. A photomultiplier tube was used to detect emission normal to the sample surface. All structural, morphological and optical measurements were performed at room temperature.

## RESULTS AND DISCUSSION

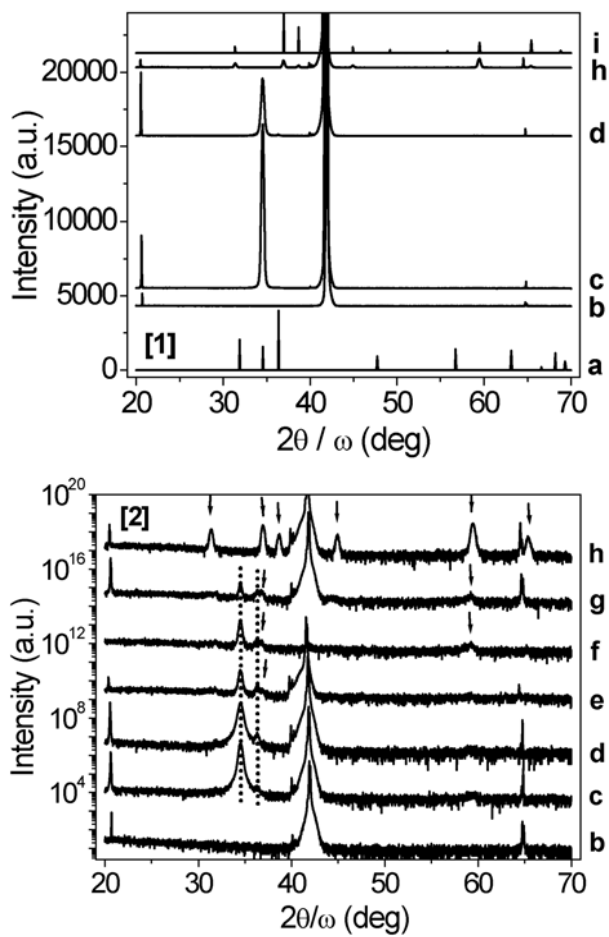

FIG. 1. Linear [1] and logarithmic scale [2] $2\theta/\omega$ XRD patterns: a) zincite ZnO reference [18], b) sapphire (0001) substrate, c) undoped, d) $Co^{2+}$, e) $Fe^{3+}$, f) $Mn^{2+}$, g) $Mn^{3+}$ and h) $V^{3+}$ doped films and i) gahnite $ZnAl_2O_4$ reference [18]. ZnO and $ZnAl_2O_4$ peaks are marked with dotted lines and arrows, respectively. Scans are arbitrarily offset for clarity.

The diffraction patterns in figure 1 show the crystallographic phases present in the samples. Each film shows the $K\alpha_1$ and $K\alpha_2$ sapphire 0006 peaks around 42 ° and other sapphire peaks at 20.70 °, 64.72 ° and 64.90 °.

The undoped and $Co^{2+}$ doped films are single phase ZnO with a strong 0002 peak at 34.54 °. A high degree of preferential orientation is evident, giving rise to spectra resembling single crystal diffraction patterns, although the films are polycrystalline. The c-axis of most ZnO crystallites aligns with the c-axis of the substrate in the [0001] direction. The crystals therefore stand with c-axis perpendicular to the substrate surface. The logarithmic scale (fig. 1[2]) reveals a second peak at 36.34 ° for all ZnO samples; this corresponds to the $10\bar{1}1$ reflection, which is the dominant peak in randomly orientated polycrystalline ZnO (fig. 1a).

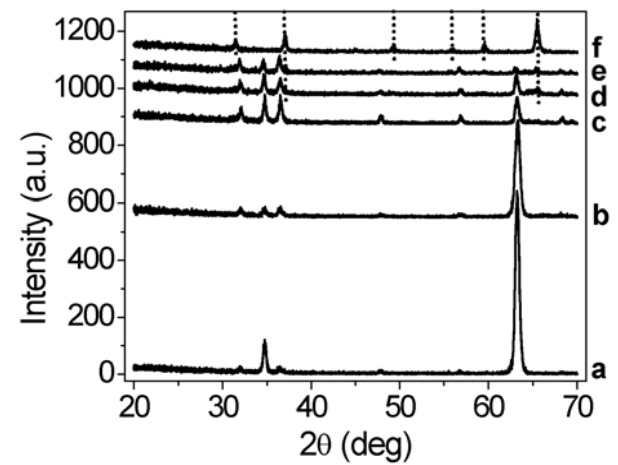

FIG. 2. Grazing incidence $2\theta$ XRD patterns at $\omega \approx 0.35$ °: a) undoped, b) $Co^{2+}$, c) $Fe^{3+}$, d) $Mn^{2+}$, e) $Mn^{3+}$ and f) $V^{3+}$ doped films. $ZnAl_2O_4$ peaks are marked with dotted lines; all other peaks originate from ZnO. Scans are arbitrarily offset for clarity.

A mixture of hexagonal ZnO and cubic (spinel) $ZnAl_2O_4$ phases is evident in $Fe^{3+}$, $Mn^{2+}$ and $Mn^{3+}$ doped films. The spinel is the product of the reaction between ZnO and sapphire: $ZnO + Al_2O_3 = ZnAl_2O_4$ [12]. Relative 0002 peak intensities show that the quality of these films is low compared to undoped and $Co^{2+}$ doped samples. The sapphire peaks are almost absent in the $Mn^{2+}$ scan (fig. 1f) because the c-axes of sapphire and ZnO are misaligned in this sample. Only spinel ZnAl$_2$O$_4$ peaks are visible above the noise level for the V$^{3+}$ doped film (scan 1h), suggesting extensive intermixing of the initial film with the substrate. The ZnAl$_2$O$_4$ phase is polycrystalline, with a slightly preferred orientation in the [111] direction since the main peak is at 59.49 ° (333 reflection) as opposed to 36.93 ° (311 reflection) in a randomly orientated powder (fig. 1i).

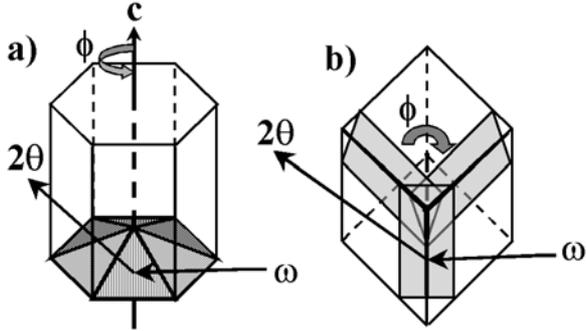

FIG. 3. Schematic diagrams of hexagonal 6-fold symmetry about [0001], and cubic 3-fold symmetry about [111], respectively in a) zincite unit cell, (10$\bar{1}$3) and commensurate planes are shaded, and b) gahnite unit cell, (440) and commensurate planes are shaded. The grazing incidence beam geometry and axis for φ rotation are illustrated.

A grazing incidence configuration, typically with ω fixed at 0.35 ° with respect to a horizontal beam provided access to specific planes in the ZnO and ZnAl$_2$O$_4$ crystallites. Contributions from the substrate are also excluded at such low angles of incidence. Figure 2 shows all of the polycrystalline ZnO reflections since the sample is tilted, giving access to more internal planes. A prominent ZnO 10$\bar{1}$3 peak is present at 63.20 ° in the undoped and Co$^{2+}$ doped samples. The intensity of the 10$\bar{1}$3 peak relative to the other reflections diminishes passing from the undoped to the Mn$^{3+}$ doped samples (scans 2a-e). The ZnAl$_2$O$_4$ phase is not apparent in the Fe$^{3+}$ doped scan (fig. 2c), but two oxide phases are visible in both Mn doped films (scans 2d-e), suggesting that the reaction had not spread beyond the interfacial region in the Fe$^{3+}$ doped sample. The V$^{3+}$ doped film again shows only the spinel phase, with a dominant 440 reflection at 65.47 ° (scan 2f).

Whilst at grazing incidence, 2θ was fixed either at the ZnO 10$\bar{1}$3 peak or the ZnAl$_2$O$_4$ 440 peak, then the sample was rotated 360 ° to perform a φ scan in the azimuthal circle in order to explore the hexagonal 6-fold and cubic 3-fold symmetry in ZnO (figure 3a) and ZnAl$_2$O$_4$ (figure 3b), respectively.

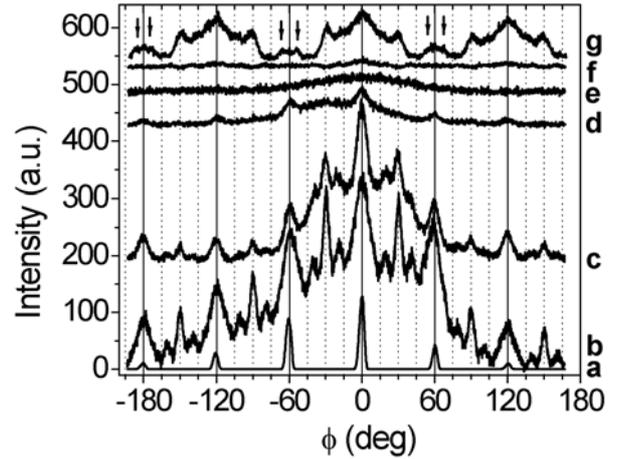

FIG. 4. Grazing incidence XRD φ scans about: Hexagonal wurtzite 10$\bar{1}$3 peak of a) GaN (0002) reference film, b) undoped, c) Co$^{2+}$, d) Fe$^{3+}$ and e) Mn$^{2+}$ doped films, and cubic gahnite 440 peak of f) Mn$^{2+}$ doped and g) V$^{3+}$ doped film. Peaks marked with arrows are referred to in the text. Scans are arbitrarily offset for clarity.

Figure 1 demonstrates the high degree of c-axis orientation, and figure 4 shows that the crystallites are also highly directional in the (a,b) plane. None of the films are as ordered as the CVD-grown GaN reference single crystal (Aixtron AG, Germany), which has a very similar wurtzite structure to ZnO, but it is clear that the ZnO crystallites are not randomly arranged in the (a,b) plane. The 60 ° separated peaks in GaN (scan 4a) are reflections from six commensurate planes in the hexagonal lattice shown in figure 3a.

The primary peaks in the undoped, Co$^{2+}$ and Fe$^{3+}$ doped samples have a 60 ° separation, showing that most of the hexagonal crystallites have the same orientation in the (a,b) plane. However, some of the undoped and Co$^{2+}$ doped crystallites are rotated 30 ° relative to most of the crystallites, giving rise to a set of peaks separated by 60 ° and offset from the primary orientation by 30

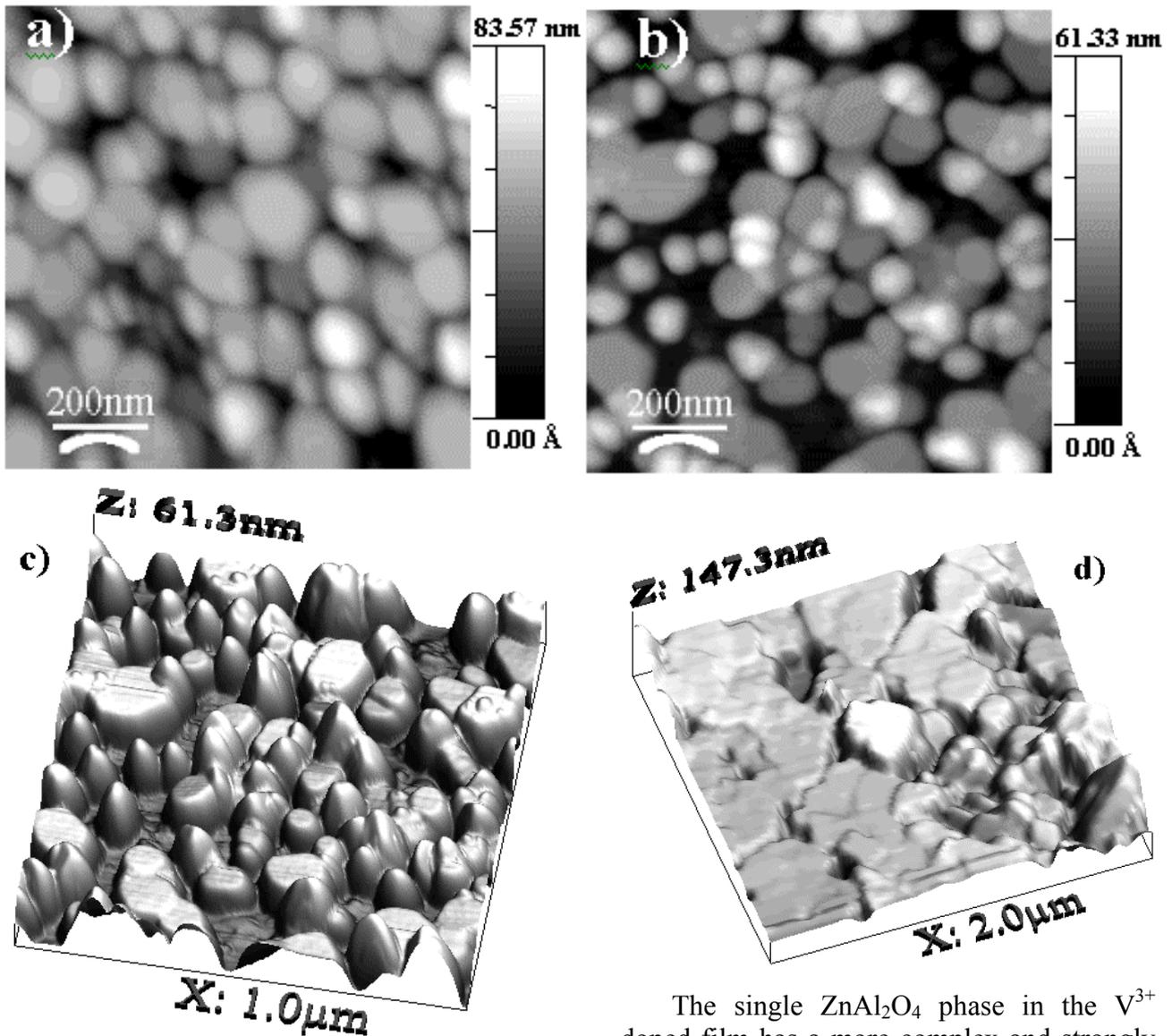

FIG. 5. 2D and 3D AFM images of: a) undoped, b-c) $Co^{2+}$ doped and d) $V^{3+}$ doped films.

°. Two more orientations are evident in the undoped and $Co^{2+}$ doped sample, one offset by 20 ° and the other by 40 ° relative to the primary direction. A degree of randomorientation is evident in all of the films, as shown by the background signal. The broad featureless curve from the $Mn^{2+}$ doped film (scan 4e) shows that no preferential orientation was found in the hexagonal phase of the film. However, a slight preference for specific orientations is visible in the cubic phase (scan 4f), showing a primary set of three small features separated by 120 °.

The single $ZnAl_2O_4$ phase in the $V^{3+}$ doped film has a more complex and strongly developed structure than the $Mn^{2+}$ doped sample. Within the 120 ° separation of the three primary peaks, sets of 120 °-separated peaks are offset from the primary peaks by 30, 60 and 90 °. Additionally, two satellite peaks are visible, shouldering the peaks offset from the primary set by 60 ° (arrows on scan 4g). These are offset from the 60 ° peaks by approximately - and + 6 °, and are therefore offset from the main peaks by about 54 and 66 °. So in total six distinct orientations are observed in the $V^{3+}$ doped sample. This complex arrangement of directions is repeatable over each 120 ° cycle separating the primary peaks. The $Mn^{3+}$ doped sample was not crystalline enough to display preferential orientation in either the hexagonal or cubic phases.

AFM yielded additional insight into the orientation of crystallites in the (a,b) plane. Figure 5a shows that the undoped film consists mainly of closely packed crystallites with a few small voids. The close proximity of the grains makes the shape and orientation of the crystallites difficult to distinguish and the size difficult to measure from the image. However, an approximate particle size of 185 nm was calculated using the Scherrer formula. The $Co^{2+}$ doped film has a more open structure with large voids that are irregularly shaped and randomly distributed. The discontinuity of the $Co^{2+}$ doped film made the size and arrangement of crystallites easier to determine. The diameter of the crystallites was measured from profiles of image 5b as the trough-to-trough horizontal distance across each grain. Figure 6 shows that the $Co^{2+}$ doped ZnO crystallites are approximately 100 nm across at the base. This value is close to the particle size estimated using the Scherrer formula (line on fig. 6). Figures 5b-c show that $Co^{2+}$ doped grains have hexagonal bases, and the flat edges of many hexagonal crystallites appear to be aligned in the (a,b) plane.

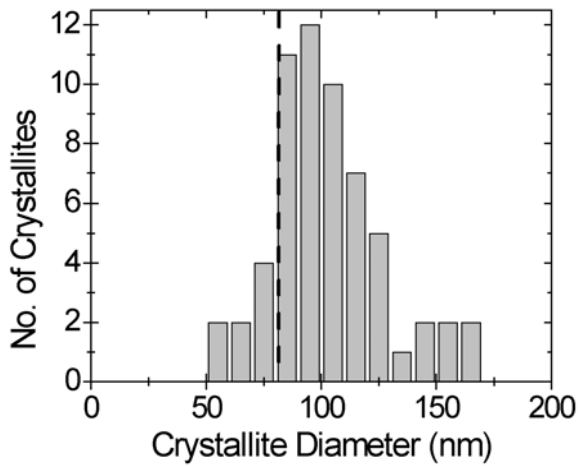

FIG. 6. Approximate base diameter of $Co^{2+}$ doped ZnO crystallites, measured from AFM image 5b. Dashed line indicates average $Co^{2+}$ doped ZnO particle size derived using the Scherrer formula.

Since the cantilever tip was 15-20 μm high and < 20 nm across at the point, it was able to follow the profile of the film, through the voids to the substrate, resulting in thickness a measurement of about 40±10 nm for all films. Undoped and $Co^{2+}$ doped grains have different height distributions. In the undoped film, all the grains are approximately the same height. However, some $Co^{2+}$ doped crystallites are low flattened hexagons about 30-40 nm high, while many others are taller and taper to a point at around 40-60 nm.

The $V^{3+}$ doped film is very different to the others. The film is densely packed and has large regions consisting either of slanted blocks or small, rough, irregular grains. The large blocks are approximately 400 nm across and appear to be similarly orientated. The $Mn^{2+}$ doped film displays a mixture of hexagonal and cubic ordering (image not shown).

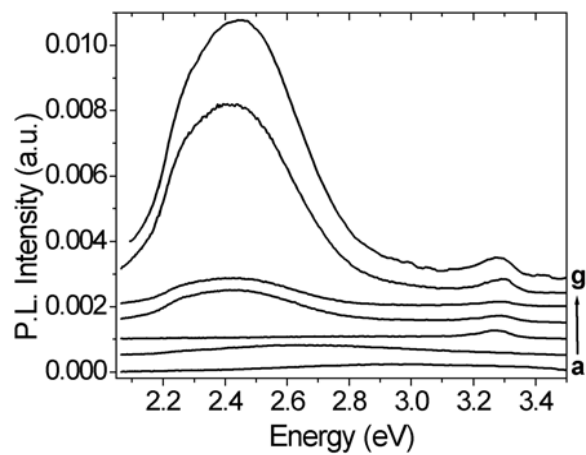

FIG. 7. Room temperature photoluminescence spectra of: a) sapphire substrate, b) $Co^{2+}$, c) $V^{3+}$, d) $Mn^{3+}$, e) $Mn^{2+}$, f) $Fe^{3+}$ and g) undoped films. Excitation wavelength 325 nm. Scans are arbitrarily offset for clarity.

Due to the low volume of the $Co^{2+}$ doped sample, photoluminescence hardly exceeded the background level (scan 7b). The characteristic green mid-band (2.4 eV) and UV band-edge (3.3 eV) emissions of ZnO [5,19] are displayed by undoped, $Fe^{3+}$, $Mn^{2+}$ and $Mn^{3+}$ films (scans 7g-d, respectively). The broad deep-level emission band originates from structural defects, which form electron-hole recombination centres. The band-edge luminescence is due to free exciton emission or exciton transitions with shallow donors or acceptors; this feature is usually associated with high structural quality. Contrary to this interpretation, the trend observed in figure 7 demonstrates that the band-edge to mid-band intensity ratio increases as ZnO structural quality diminishes, culminating with zero ZnO mid-band emission from the $V^{3+}$ doped sample (scan 7c). Figures 1, 2 and 4 show that the reduction in ZnO structural quality is accompanied by an increase in the proportion of $ZnAl_2O_4$ present, suggesting that this phase may be responsible for the observed trend. Since $ZnAl_2O_4$ has a direct band gap of 4.0 – 4.5 eV [20], the excitation energy of 3.8 eV is insufficient to initiate absorption across the energy gap, thereby diminishing the luminescent intensity in the range measured. Emission from the $V^{3+}$ doped sample coincides with the ZnO band edge, so could originate from residual ZnO below the detectable limit for XRD. Note that the band-edge to mid-band intensity ratio is the same for $Mn^{2+}$ and $Mn^{3+}$ doped samples.

## CONCLUSIONS

A novel sol-gel synthesis of c-axis orientated undoped and transition metal doped ZnO films was described. X-ray diffraction revealed single-phase ZnO for undoped and Co doped films. Vanadium doped films formed $ZnAl_2O_4$, which resulted from the reaction between ZnO and the sapphire substrate. This reaction was probably catalysed by the vanadium ions. Mixed hexagonal zincite ZnO and cubic spinel $ZnAl_2O_4$ phases were detected in the iron and manganese doped samples. The highly directional nature of the ZnO crystallites was demonstrated by XRD. A high degree of long-range structural order was evident both parallel and perpendicular to the substrate surface for undoped and Co doped films. V doped $ZnAl_2O_4$ crystallites were also preferentially orientated. The ordered arrangement of ZnO and $ZnAl_2O_4$ films was confirmed by AFM imaging, which showed that undoped and Co doped films had different distributions of crystallites, and the V doped $ZnAl_2O_4$ film contained regions of orientated crystallites. Optical characterisation demonstrated the high structural quality of the films by the presence of the ZnO band edge feature. Mid-band emission appeared to be suppressed by the presence of $ZnAl_2O_4$.

Our results illustrate the utility of sol-gel methods for producing high quality oxide semiconductors suitable for a wide range of technological applications.

## ACKNOWLEDGEMENTS

The authors thank the University of Nottingham Institute for Materials Technology (UNIMAT) for funding. NRSF also thanks Prof. Tom Foxon for helpful discussions, Dr Philip Moriarty for use of AFM equipment and Dr I. Harrison for use of PL instruments.